\documentstyle[aps,graphicx,multicol]{revtex}
\newcommand\be{\begin{eqnarray}}
\newcommand\ee{\end{eqnarray}}
\newcommand\ba{\begin{array}}
\newcommand\ea{\end{array}}
\def\r{\rangle}
\def\l{\langle}
\def\T{{\rm Tr}}

\def\cI{{\cal I}}

\def\cE{{\cal E}}

\def\cA{{\cal A}}
\def\openone{{\it I}}

\begin{document}
\title{Process reconstruction from incomplete and/or inconsistent data}
\author{M\'ario Ziman$^{1,2}$, Martin Plesch$^{1}$, Vladim\'\i r Bu\v zek$^{1,2}$ }
\address{
$^1$Research Center for Quantum Information,
Slovak Academy of Sciences,
D\'ubravsk\'a cesta 9, 845 11 Bratislava, Slovakia\\
$^2$Faculty of Informatics, Masaryk University, Botanick\'a 68a,
602 00 Brno, Czech Republic
}
\maketitle
\begin{abstract}
 We analyze how an action of a qubit channel (map) can
be estimated from the measured data that are incomplete or even
inconsistent. That is, we consider situations when measurement
statistics is insufficient to determine consistent probability
distributions. As a consequence either the estimation
(reconstruction) of the channel completely fails or it results in
an unphysical channel (i.e., the corresponding map is not
completely positive). We present a regularization procedure that
allows us to derive physically reasonable estimates (approximations) of quantum
channels. We illustrate our procedure on specific examples and we
show that the procedure can be also used for a derivation of
optimal approximations of operations that are forbidden by the
laws of quantum mechanics (e.g., the universal NOT gate).
\end{abstract}

%
%

\begin{multicols}{2}

\section{Introduction}

For any reliable quantum information processing it is important to know how states of
quantum systems are transformed under the action of quantum channels (maps). It is therefore
essential to develop tools by means of which we can acquire knowledge about properties of
quantum channels. Providing we have no prior knowledge about an action of a particular transformation
our task is to determine characteristics of a corresponding quantum map based on correlations between
input and out states of quantum systems that serve as probes of the channel.

In principle, the action of a quantum channel can be probed in two different
ways: {\bf (1)} The first option is to use as an input a {\it single} entangled state of a bi-partite system
\cite{dariano,dariano2,jezek,hradil}. One particle (e.g., a qubit) of this bi-partite system is transformed
under the action of the channel while the second particle remains unchanged (or evolves according to
a known transformation). By performing a complete quantum tomography of the bi-partite system at the output of the channel
and comparing the input and output states one can determine what is a specific action of the channel
under consideration. {\bf (2)} The second option is to use a
collection of linearly independent test states (forming a basis
of the vector space of all hermitian operators) \cite{poyatos,Chuang1997,Buzek1998,nielsen}.
By performing a correlation measurement between a specific input and corresponding output state (that has
been tomographically reconstructed) we can determine the map that characterize the quantum channel. For a total
determination of the map we have to use a complete set of test states.

Reconstruction of quantum channels using entangled states might seem to be more efficient since a preparation of just
one state of an entangled pair is required. Nevertheless, there are two technical problems that make this approach less practical
than a utilization of single particle states. Specifically, one has to generate input bi-partite entangled states with a very high
fidelity (it is essential that for a reliable channel reconstruction the input test states have to be prepared with a very high fidelity).
Simultaneously, the tomographic reconstruction of a bipartite entangled state at the output of the channel has to be almost perfect.
But the most difficult
obstacle is to secure that the ``reference'' particle from the entangled pair does not undergo uncontrolled changes during the
time when the second particle is affected by the action of the quantum channel.
These conditions are rather difficult to met.
For this reason in the present paper we will concentrate
our attention on the second scenario. This approach  is based on the fact that the action of any channel is described by a linear
map $\cE$ and therefore it is completely determined by its action on basis elements, i.e.
a set of linearly independent states, which play the role of
{\it test states}. We assume, that these single-partite test states are known (i.e., their preparation is under a
complete control). Thus the process of the channel
reconstruction reduces to the re\-con\-struc\-tions of single-particle states at the output of the quantum channel.
The number of test states equals $d^2$, where $d$ is the dimension of
the Hilbert space of a quantum system under consideration. In order to fully characterize the action of the quantum
channel acting on a such quantum system we need  $d^2(d^2-1)$ real parameters, i.e. in the case of a qubit channel we need 12 real parameters.

In our previous paper \cite{ziman} we have analyzed the question of the process
reconstruction from incomplete experimental data. That is, we have considered a situation when the correlation
input-output measurements do not allow for a unique determination of the $d^2(d^2-1)$ real parameters. We have shown
how to proceed with the process estimation in the case of incomplete experimental data. In this paper we want to
study more delicate problem - how to perform process estimation when the experimental data are incomplete and/or
they are not consistent. That is, the straightforward estimation leads to maps that are not physical (i.e., they are positive
but not completely positive) \cite{wunderlich}. We will present a regularization procedure that allows us to handle such situations.
Moreover, we will show that this procedure also allow us to determine optimal approximations of non-physical operations.
As an example we will analyze in detail the so called universal NOT gate. The paper is organized as follows. In Section 2 we briefly
describe properties of qubit state space while in Section 3 we discuss the structure of space of qubit operations (channels).
Sections 4 and 5 are  devoted to the reconstruction of quantum channels from incomplete data. Approximations of non-physical operations
are derived in Section 6. The conclusions are presented in Section 7.

\section{Structure of qubit state space}
Firstly, let us consider a simple geometrical representation of a qubit
state space. This state space has a topology of a sphere that is often called as the Bloch sphere.
The set of all hermitian operators form a real
vector space endowed
with a scalar product defined by the relation $(A|B)=\T A B$. Consequently, any
operator can be written as a linear combination
of operators that form an orthogonal basis.
The set of the Pauli $\sigma$-matrices, i.e. $\{\openone,\sigma_x,\sigma_y,\sigma_z\}\equiv \{\openone,\vec{\sigma}\}$,
represents a standard choice of the
operator basis for a qubit (see, e.g., Ref.~\cite{nielsen}). The state space of a qubit
is a subset of all hermitian operators with a unit trace.
Except the operator $\sigma_0=\openone$
all other members of the $\sigma$-basis are traceless. Therefore
any operator with a unit
trace can be written as
$\varrho=\frac{1}{2}(\openone+\vec{r}\cdot\vec{\sigma})$.
Such operators represent a quantum
state only when the operator
$\varrho$ is positive, i.e. $|\vec{r}|\le 1$. In this way we
obtain the Bloch sphere representation of qubit states,
$\varrho\leftrightarrow\vec{r}$.
A state reconstruction is a task of experimental
specification of the vector $\vec{r}=(x,y,z)$.
>From the orthogonality condition  $\T\sigma_k\sigma_l=2\delta_{kl}$ we find that
components of the vector $\vec{r}$
are determined by an expression $\vec{r}_k=\T\varrho\vec{\sigma}_k$. That is, they are equal
to mean values of the hermitian operators (measurements)
$\sigma_x,\sigma_y,\sigma_z$. So the complete reconstruction
is straightforward: All one has to do is to measure  mean values of
 three system operators $\sigma_x,\sigma_y,\sigma_z$.

Let us note that sometimes the reconstructed density operator
may not satisfy the condition $|\vec{r}|\le 1$. This failure of the
reconstruction scheme is usually caused by an inconsistent measurement
statistics which results in an incorrect identification of probabilities
and consequently in the derivation of false mean values.
The easiest way how to perform a regularization of the
reconstruction in this case is the following one:
The reconstructed density operator has to have always a unit trace, i.e. the operator is
represented by a vector $\vec{r}$ though it might have a length larger than unity. In this case
the reconstruction procedure fails since the estimated operator is not physical.
One can argue that an actual physical
state is the closest one to the ``reconstructed'' operator represented by a
point on the Bloch sphere (a pure state) with $\vec{r}_c$
pointing in the same direction as the reconstructed vector $\vec{r}$. Formally the regularization
corresponds to a multiplication of the original vector $\vec{r}$
by  a positive constant $k$, i.e. $\vec{r}_c=k\vec{r}$. From the physical
point of view this regularization can be understood  as an admixture of a ``white'' noise described by the operator
$\frac{1}{2}\openone$ [that is represented by the center of the Bloch sphere, i.e.
$\vec{0}=(0,0,0)$] to the measured data. Formally this ``regularization'' procedure reads
\be
\varrho_c = k\varrho+(1-k)\frac{1}{2}\openone =
\frac{1}{2}(\openone+k\vec{r}\cdot\vec{\sigma})\, .
\ee
Such correction corresponds to the addition of completely
random and equally distributed events (``clicks'') to the outcome statistics of measurement results. In what follows we will
utilize an analogue of this intuitive picture to regularize reconstructions of maps describing quantum channels.

\section{Structure of qubit channels}
The structure of qubit channels is known mainly due to work of M.B.Ruskai
et al.\cite{ruskai}. Let us  briefly summarize main properties of qubit channels.
Any completely positive map $\cE$ can be imagined as an affine
transformation of the vector $\vec{r}$, i.e.
$\vec{r}\to\vec{r}^\prime= T\vec{r}+\vec{t}$, where
$T$ is a real 3x3 matrix and $\vec{t}$ is a translation. However,
this form guarantees only the preservation of a trace and the hermiticity  of the
transformation $\cE$. In fact, the set of all completely
positive tracepreserving maps
forms a specific convex subset of all affine
transformations. For qubits the number of parameters specifying
the channel equals to 12.  Because of the affinity of any
evolution map $\cE$, one can use the following matrix representation
\be
\cE=\left(
\begin{array}{cccc}
1 & \vec{0} \\
\vec{t} & T
\end{array}
\right)\, , \ \ {\rm and}\ \
\varrho=\left(
\begin{array}{c}
1\\
\vec{r}
\end{array}
\right)\, .
\ee
The coefficients of the matrix $\cE$ are given by a relation
$\cE_{kl}=\T(\sigma_k\cE[\sigma_l])$, where $\sigma_{k(l)}$ are Pauli
$\sigma$ matrices.

Any matrix $T$ can be written in the so-called {\it singular value decomposition}, i.e.
$T=R_UDR_V$ where $R_U,R_V$ are orthogonal rotations and
$D={\rm diag}\{\lambda_1,\lambda_2,\lambda_3\}$ is a diagonal matrix
with $\lambda_k$ being the singular values of $T$.
Each three-dimensional
orthogonal rotation $R_U$ (element of the group S0(3)) is related
to some qubit unitary transformation $U$ (an element of the group SU(2))
via the relation $U\varrho U^\dagger = \frac{1}{2}(\openone+(R_U\vec{r})\cdot\vec{\sigma})$.
This means that any
map $\cE$ is a member of less-parametric family of maps
of the ``diagonal form'' $\Phi_\cE$. In particular,
$\cE[\varrho]=U\Phi_\cE[V^\dagger\varrho V]U^\dagger$ where
$U,V$ are unitary operators. This reduction of parameters is
very helpful, and most of the properties (also complete positivity)
of $\cE$ is reflected by the properties of $\Phi_\cE$.
The map $\cE$ is completely positive (CP) only if
$\Phi_\cE$ is also CP. Let us note that $\Phi_\cE$ is determined not only by
the matrix $D$, but also by a new translation vector $\vec{\tau}=R_U\vec{t}$,
i.e. under the action of the map $\Phi_\cE$ the
Bloch sphere transforms as follows $r_j\to r_j^\prime=\lambda_j r_j+\tau_j$.
\begin{figure}
\includegraphics[width=8cm]{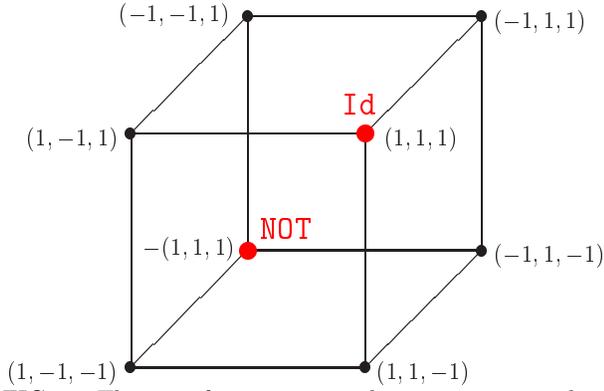}
\caption{The set of positive unital trace-preserving linear
maps $\Phi_\cE$ parametrized by three real parameters
$\lambda_1,\lambda_2,\lambda_3$. The point $\lambda_1=\lambda_2=\lambda_3=-1$
corresponds to a physically unrealizable transformation -
the universal NOT (denoted as NOT in the figure) operation 
, i.e.
The point $\lambda_1=\lambda_2=\lambda_3=1$ represents the identity map $\cI$ (denoted as Id)}
\label{kocka}
\end{figure}

\begin{figure}
\begin{center}
\includegraphics[width=6cm]{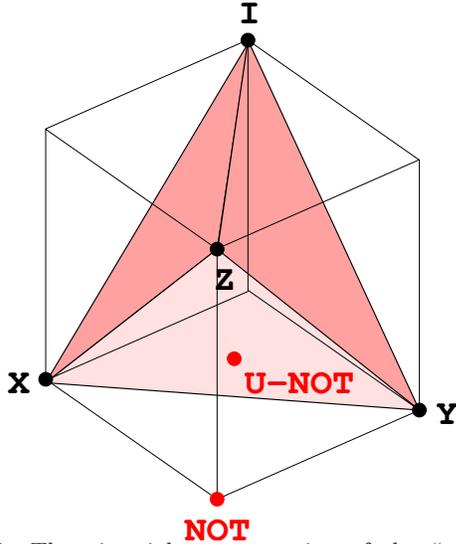}
\caption{The pictorial representation of the ``position''of  completely positive unital  maps
in the space of all positive unital maps. The CP unital maps form
a tetrahedron with four unitary transformations in its vertices
(extremal points), that correspond to $\sigma$-matrices.
The un-physical universal
NOT operation (denoted as NOT) and its best completely positive
approximation called the optimal universal NOT (denoted as U-NOT) are shown as well. The optimal universal
NOT is represented by the point $\lambda_1=\lambda_2=\lambda_3=-1/3$.}
\label{unital_cp}
\end{center}
\end{figure}

A special type of completely positive maps is a class of {\it unital} transformations, i.e.  maps for
which the total mixture (the center of the Bloch sphere) is not affected by the transformation.
It means that
the translation term vanishes, i.e $\vec{t}=\vec{\tau}=\vec{0}$. In this case the
geometrical analysis is quite simple. The positivity of
the transformation $\Phi_\cE$
corresponds to the conditions $|\lambda_k|\le 1$, i.e. these transformations are represented by points lying
inside a cube (see Fig.\ref{kocka}). The conditions
of the complete positivity \cite{ruskai} requires the validity of the
following four inequalities
\be
|\lambda_1\pm\lambda_2|\le|1\pm\lambda_3| \, .
\label{cp}
\ee
These inequalities specify a tetrahedron inside a cube
of all positive unital maps with the extreme points being four unitary
transformations $\openone,\sigma_x,\sigma_y,\sigma_z$
(see Fig.\ref{unital_cp}).
As a result of this analysis one can conclude that the unital
completely positive maps $\Phi_\cE$ form
a tetrahedron.

\section{The channel reconstruction: Inconsistent statistics}
Let us consider a situation when we
want to estimate a channel, but neither experimental data nor non-trivial prior knowledge are available.
Following the approach utilized for quantum state estimation \cite{Buzek2001} one can assume that in the
absence of knowledge about the character of quantum channel the most reliable estimation of this channel
corresponds to an equally weighted ``average'' over all possible quantum maps.
The question is what is the average over all completely
positive maps? We have already argued \cite{ziman} that for qubits this average
is the map $\cA$ that transforms the whole state space into
the total mixture, i.e. $\cA[\varrho]=\frac{1}{2}\openone$. The reasoning goes as follows: The maps
$\cE_{\pm}\leftrightarrow (T,\pm\vec{t})$ have the property that if one of them is completely positive
the second one is CP as well. Consequently, the average
$\frac{1}{2}(\cE_+ +\cE_-)$ is a unital map, i.e. the average of all
maps will be a unital map. As we already said, the unital maps
(up to unitary transformations) form a tetrahedron. The average over
all points in tetrahedron is represented by the center of the tetrahedron, i.e.
$\lambda_1=\lambda_2=\lambda_3=0$, which corresponds to
the contraction into the total mixture. Therefore, the
average over all completely positive maps of acting on a qubit is the contraction
of the whole Bloch sphere into its center,
i.e. $\cA[\varrho]=\frac{1}{2}\openone$.

The result of a complete reconstruction based on the
four test states (i.e., any collection of four
mutually linearly independent qubit states) is a map
that is for sure trace-preserving and positive.
However, it could happen that it is not completely positive.
How to extract the physical map $\cE_c$ from an unphysical result $\cE$?
One way is to follow a similar reasoning like for the state reconstruction,
when an unphysical result was corrected (regularized) by adding a noise into the system.

When the reconstructed map is not completely positive we can regularize this result by
an analogue of the total mixture, i.e. the map $\cA$.
In particular, this regularization of a quantum channel reads
\be
\cE_c=k\cE+(1-k)\cA=\left(
\ba{cc}
1 & \vec{0} \\
k\vec{t} & kT
\ea
\right)\, .
\ee
The correction (regularization) corresponds to a ``minimal'' adjustment of the parameter $k$ such that
the map $\cE_c$ is completely positive.

Let us estimate what is this minimal (critical) value of the parameter
$k$, i.e. the value which surely regularizes any positive map and transforms it into a CP map.
Trivially, it is enough to set $k=0$. In this case we completely
ignore the measured data and the ``corrected'' map is $\cA$.
However, we are interested in some nontrivial bound, i.e. in the largest
possible value of $k$. As we have already mentioned, the reconstructed map is always
positive. Consider, for simplicity, that the map is also unital.
Then the ``worst'' example of a positive map, which is not completely positive,
is the universal NOT operation. In this case the distance between this
map and the tetrahedron of completely positive maps is extremal
(see Fig.\ref{unital_cp}).
This artificial example serves as a good test of our method,
and gives us some bound on $k$, i.e. a value that surely corrects each result.
The conditions of the complete positivity given in Eq.(\ref{cp})
imply that $k=1/3$, i.e. $\lambda_1=\lambda_2=\lambda_3=-1/3$
(see Fig.\ref{unital_cp}). Surprisingly, this is the same result
as the one that has been obtained in Ref.~\cite{werner} where the best (optimal)
completely positive approximation of the universal NOT operation, i.e.
an optimal universal NOT machine has been presented. In this sense our correction
method works optimally.

\section{The channel reconstruction: incomplete data}
In this section we will present the reconstruction scheme
which can be used when the
number of test states is reduced so that the complete reconstruction of the
channel cannot be performed \cite{ziman}. On the other hand we assume that each
of the state used is represented by an infinite ensemble of identically prepared
states so the complete tomography of a corresponding state at the output of the channel
can be performed.

In the case of qubit channels the aim is to perform
the reconstruction based on $n=0,1,2,3,4$ measured input-output correlations of the form
$\varrho_j\to\varrho_j^\prime$. We have already shown that
having no information ($n=0$), the best estimation of the map is the contraction
into a total mixture (an average over all quantum channels)
i.e. $\cE_0=\cA$. Motivated by such result, the reconstruction strategy (see
 Ref.~\cite{ziman}) is as follows: All undetermined states (belonging to
the complement of the linear span of the used input states)
are assumed to be transformed into the total mixture. That is, if a given test state is not
explicitly used for a channel reconstruction (i.e., it is not know how this state
is transformed by the action of the channel) it is {\it assumed} that the channel transforms this
state into a total mixture. This additional assumption complements the knowledge of how
other test states are actually transformed by the channel and allows us to use the deterministic procedure
of channel reconstruction (for details see e.g. Refs.~\cite{Chuang1997,poyatos,Buzek1998}).

Because of the {\it ad hoc} assumption about the transformation of unused test states it
might happen that the resulting map
is not completely positive. In this case the estimation procedure has to be
complemented
by a search for a map for which the total mixture is ``shifted'' as
little as possible, i.e. the estimated channel is preferably unital.
A specific situation occurs when the data
contain information about the transformation of the total mixture.
In this case the strategy suggests to use the state
$\cE[\frac{1}{2}\openone]$ as the state on which all other
undetermined states are mapped. As before, if such map is not
completely positive we have to search for a map, for which the deviation from
the transformation $\cE[\frac{1}{2}\openone]$ is minimal.

The method described above is discussed in detail in Ref.~\cite{ziman}.
We note that this (incomplete) reconstruction can also fail (it gives no result), because
it can happen that no physical channel is
is compatible with given experimental data. Even if the reconstructed
operators $\varrho_j^\prime$ describe valid quantum states, the
incomplete data can be in a contradiction with the condition of a complete
positivity. In what follows we will briefly
describe our strategy
on a particular example - estimation of the identity channel (i.e. the channel, that
does not change input states at all).
For more detailed and more general description of this strategy
see Ref.~\cite{ziman}.

\subsection{Case study: identity channel}
In what follows we will perform a step-by-step reconstruction of a qubit channel
based on a knowledge of how a single, two and three test states are transformed under
the action of a given channel.

{\bf Single test state.} Let us assume that our knowledge about the action
of a particular channel is represented by the assignment
\be
\varrho_1\to\varrho_1^\prime=\varrho_1\, .
\ee
We remind us that
each state can be written in the form
$\varrho_1=\frac{1}{2}(\openone+\vec{r}\cdot\vec{\sigma})=
\frac{1}{2}(\openone+ w S_z)$, where $S_z=|\psi\r\l\psi|-
|\psi_\perp\r\l\psi_\perp|$ with $|\psi\r,|\psi_\perp\r$ being eigenvectors
of the operator $\varrho_1$ and $w$ describes an impurity of the state under consideration,
i.e. $w=\sqrt{1-2\T\varrho_1^2}$. If $w=0$ then $\varrho_1$ describes maximally mixed state
and for $w=1$ the state is pure. In the Bloch-sphere picture the parameter
$w$ corresponds to the distance between the total mixture (the center of the sphere) and a point
corresponding to the given state. One can define a new
operator basis $S_x,S_y,S_z$ such that $S_z$ is given as before, and
$S_j= U \sigma_j U^\dagger$ with unitary $U$. In this new basis
the action of the channel is described by the matrix
\be
\cE_1=\left(
\ba{cccc}
1 & 0 & 0 & 0 \\
x & a & d & 0 \\
y & b & e & 0 \\
z & c & f & 1
\ea
\right)
\ee
and the task of the estimation is to specify all  matrix elements.
Our strategy suggests  that all states (belonging to the complement
of the linear span of $\varrho_1$) are transformed into the total mixture, i.e. we
set all the parameters equal to zero.

\begin{figure}
\includegraphics[width=8cm]{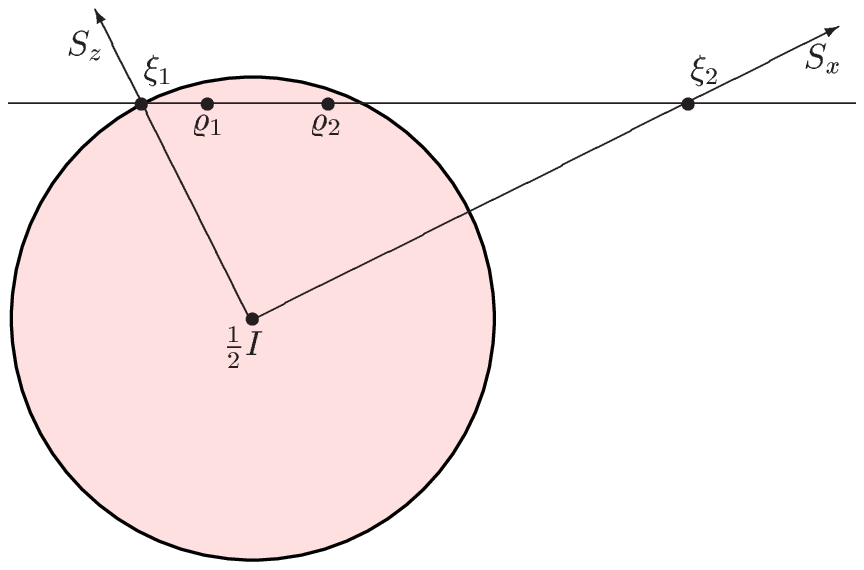}
\caption{The pictorial description of the channel estimation when two states $\rho_1$ and $\rho_2$
are used to test the action of a quantum channel and they are transformed according to Eq.~(\ref{6}).  }
\label{2states}
\end{figure}

{\bf Two test states.} In this case the knowledge about the action of the channel is
represented by a transformation of two test state given
by the following relations
\be
\label{6}
\varrho_1\to\varrho_1^\prime=\varrho_1 \, ;\ \ \ \ \ \ \ \ \ \ \ \
\varrho_2\to\varrho_2^\prime=\varrho_2\, .
\ee
Irrespective of states $\varrho_1$ and $\varrho_2$ we use, they specify a line crossing the Bloch sphere.
In particular, from Eq.~(\ref{6}) we obtain a knowledge about the transformation of all
states of the form
\be
\varrho_\lambda=\lambda\varrho_1+(1-\lambda)\varrho_2
\ee
with a real parameter $\lambda$. This line crosses the Bloch sphere in
two extremal points representing pure states. Let us choose one of these
pure states [denoted by $\xi_1=\frac{1}{2}(\openone+w S_z)$].
We also define a new operator (not necessarily positive)
$\xi_2=\frac{1}{2}(\openone+v S_x)$, which in the Bloch-sphere picture
corresponds to the intersection of the line $\varrho_\lambda$ and the
line orthogonal at the center of the Bloch sphere to the line given by points $\{ \xi_1,\frac{1}{2}\openone\}$
(see Fig.\ref{2states}).
This new state $\xi_2$ defines the operator $S_x$. Using the
new operator basis $S_x,S_y,S_z$ we can express the map characterizing the action of the channel as
\be
\cE_2=\left(
\ba{cccc}
1 & 0 & 0 & 0 \\
x & 1 & a & 0 \\
y & 0 & b & 0 \\
z & 0 & c & 1
\ea
\right)\, .
\ee
We again set
all free (unspecified by the measurement) parameters to zero.
As a result we find  that the transformation
$\cE_2={\rm diag}\{1,1,0,1\}$ is not completely positive, i.e.
it does not describe a valid quantum channel. To preserve the unitality
of the channel ($\cE_2[\frac{1}{2}\openone]=\frac{1}{2}\openone$)
one has to verify the complete positivity of the
transformation $\cE_2={\rm diag}\{1,1,k,1\}$.
We find out that the only possibility is to take $k=1$. Consequently, the
identity channel is correctly estimated, but we are still not
sure, whether the map is unital. However, we have to note one specific case,
when the total mixture is among the states $\varrho_\lambda$, i.e.
$\xi_2=\frac{1}{2}\openone$. In this case
\be
\cE_2=\left(
\ba{cccc}
1 & 0 & 0 & 0 \\
0 & e & a & 0 \\
0 & f & b & 0 \\
0 & g & c & 1
\ea
\right)
\ee
and our reconstruction procedure leads us to
the channel $\cE_2={\rm diag}\{1,0,0,1\}$, i.e. the values of all
free parameters are equal to zero. Note that in this case
$\cE_2=\cE_1$.

{\bf Three test states.}
Let us consider that the linear span of $\varrho_1,\varrho_2,\varrho_3$
does not contain the total mixture. In this case the estimated transformation
takes the form
\be
\cE_3=\left(
\ba{cccc}
1 & 0 & 0 & 0 \\
x & 1 & 0 & 0 \\
y & 0 & 1 & 0 \\
z & 0 & 0 & 1
\ea
\right)\, .
\ee
The only possibility to preserve the complete positivity of $\cE_3$
is to chose $x=y=z=0$, which is completely compatible
with our strategy. Consequently, the channel
is estimated perfectly and it is described by the transformation $\cE_3={\rm diag}\{1,1,1,1\}=\cI$.
In the case, when from the measured data it follows that
$\frac{1}{2}\openone\to\frac{1}{2}\openone$, the estimation coincides with the
reconstruction with two test states, where the transformation
of the total mixture is estimated to be
$\cE[\frac{1}{2}\openone]=\frac{1}{2}\openone$. The only difference is that
in the three-state case the unitality is guaranteed by the data. The reconstruction gives
us the same result in both cases, i.e. $\cE_3=\cI$. As a result we find that
identity channels for qubits can be uniquely identified using just three test states.

Let us summarize the incomplete reconstruction of the identity channel.
The hierarchy of estimations on different levels specified by the number of test states
is as follows
\be
\cE_0&=&\cA \, ;\\
\cE_1&=&{\rm diag}\{1,0,0,1\}\, ; \\
\cE_2&=&\cE_3=\cE_4=\cI\, .
\ee
>From our previous discussion we can conclude that the identity channel can be reconstructed
using just  three test states. Given the fact, that for any unitary channel
$\cE$ the induced map $\Phi_\cE$ represents the identity channel,
i.e. $\Phi_\cE=\cI$ we can conclude that for a complete determination of a unitary channel we
need just three test states. A unitary transformation is determined
by the choice of the basis $S_x,S_y,S_z$.

\section{Combination of data imperfections: incomplete and inconsistent data}
In this section we will study how to perform a process reconstruction
when the data obtained from the measurement are incomplete as well as inconsistent.
To secure that the incomplete reconstruction does not fail
we can adopt the regularization procedure as described earlier in the paper. In particular, we
have two options: (i) either to regularize the output test states, or (ii) to regularize the estimated map itself.

The first scenario has to be used always when  the state reconstruction of some of
the output test states $\varrho_j^\prime$ fails. However, also if
all the test states are  estimated (reconstructed) correctly (i.e. they are legitimate
physical states), the complete positivity
is not guaranteed and the process reconstruction might fail. This situation can occur even for
two test states. Specifically, the assignments $\varrho_1\to\varrho_1^\prime$
and $\varrho_2\to\varrho_2^\prime$ are compatible with some
completely positive map $\cE$ if and only if
$D(\varrho_1,t\varrho_2)\ge D(\varrho_1^\prime,t\varrho_2^\prime)$
for all real positive $t$ \cite{Carlini2003}. Unfortunately,
for three and four test states no similar
result is known. Therefore the complete positivity of the process reconstruction has to be checked for each individual case
separately.

In what follows we will utilize (demonstrate) two reconstruction strategies:
\begin{description}
\item{$\bullet$ }
We will use the reconstruction from incomplete data as discussed above.
In the case of a failure (the reconstructed channel is not a CP map) a regularization of outputs of
test states is performed (noise is admixed into the output states, so that the reconstructed map
becomes CP).
\item{$\bullet$}
We will assume that an unknown map transforms all states except the basis test
states into the total mixture, i.e. we set all free parameters to
zero. If the resulting map is not completely positive, then
the map will be regularized by adding the average channel $\cA$.
\end{description}

Let us consider a particular example that allows us to demonstrate
these two methods. In our example we will use an
``artificial'' and unphysical data generated by the universal NOT
operation \cite{werner}. The logical NOT operation is defined by
relations $|0\r\to|1\r$, $|1\r\to|0\r$
in a computer basis $\{|0\r ;|1\r\}$.
 These relations do not
completely determine a quantum channel, i.e. many quantum
channels perform such transformations. For instance, the
``classical'' NOT can be viewed as the transformation of
the Bloch sphere into the line connecting North ($|1\r\l 1|$)
and South Pole ($|0\r\l 0|$), i.e.
${\tt NOT}_{c}[\varrho]=\frac{1}{2}(\openone+\l\sigma_z\r_\varrho S_z)$
with $\l\sigma_z\r_\varrho=\T\varrho\sigma_z$. There exists also
a unitary (``quantum'') realization of the logical NOT operation, i.e.
${\tt NOT}_q[\varrho]=\sigma_x\varrho\sigma_x$. However, a natural
generalization of the NOT operation is the so called {\it universal} NOT gate, which performs the
transformation $|\psi\r\to |\psi_\perp\r$ for all states $|\psi\r$. The universal NOT gate
is not completely positive (for mixed states we will use the notation
$\varrho\to\varrho^T$). The {\it optimal universal} NOT operation
is the closest physically valid map that performs approximatively
the universal NOT. In Ref.~\cite{werner} it has been shown that
the corresponding quantum channel that maximizes the average
fidelity under given constraints reads $\cE_{\tt NOT}={\rm diag}\{1,-1/3,-1/3,-1/3\}$.

\subsection{Case study: the universal NOT gate}
\begin{figure}
\begin{center}
\includegraphics[width=8cm]{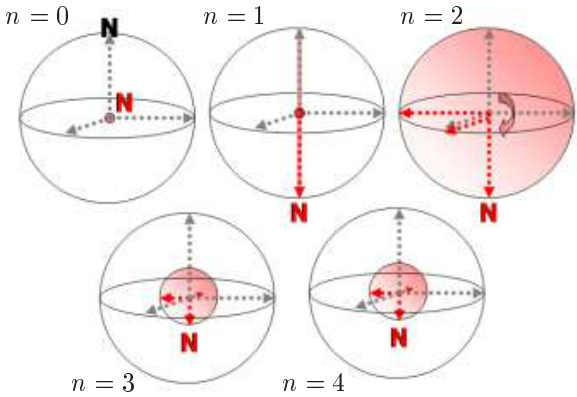}
\caption{
The figure represents how our
knowledge about the channel (the universal NOT gate)
is improved with increasing the number of test states when the first reconstruction strategy
is used.
If no measurement has been performed then the best possible
estimation of the channel is the contraction of the Bloch sphere
into its center ($n=0$). When a single test state
$\varrho_z\to\varrho_z^\prime$ is used  the optimal estimation of the transformation
is a contraction of the whole Bloch sphere into a line connecting the ``north'' (denoted as $N$) and
``south'' poles of the Bloch sphere (i.e. the points  $(0,0,\pm 1)$, respectively).
 This
contraction is represented by the figure $n=1$ in the picture.
When two test states ($n=2$) are used, i.e. $\varrho_z\to\varrho_z^\prime$
and $\varrho_y\to\varrho_y^\prime$, then the result of the estimation of the action of the quantum
channel is a rotation by the angle $\pi$  around the $x$ axis, i.e. the $\sigma_x$
operation. Finally, when three ($n=3$) or four ($n=4$) test states are
used to determine the action of the quantum channel prescribed by the universal NOT transformation, the
best estimation of the channel is the optimal universal NOT gate (see the two figures of the lower
line in the picture, both figures are the same and they describe the action of the optimal universal NOT
gate). In this case the regularization of the output test states has been used in order to estimate
a map that is completely positive. As a specific example we show how the ``north'' pole N of the Bloch sphere is transformed under
the action of  estimated maps.
}
\label{incomplete}
\end{center}
\end{figure}

Formally the universal NOT operation determines the following transformations on a set of test
(basis) states
\be
\varrho_x=\frac{1}{2}(\openone+\sigma_x)&\to&\varrho_x^\prime=
\frac{1}{2}(\openone-\sigma_x)\; ;  \\
\varrho_y=\frac{1}{2}(\openone+\sigma_y)&\to&\varrho_y^\prime=
\frac{1}{2}(\openone-\sigma_y)\; ; \\
\varrho_z=\frac{1}{2}(\openone+\sigma_z)&\to&\varrho_z^\prime=
\frac{1}{2}(\openone-\sigma_z)\; ;  \\
\varrho_{0}=\frac{1}{2}\openone&\to&\varrho_0^\prime=
\frac{1}{2}\openone \; .
\ee
Let us start with the first reconstruction strategy as described above.
In the case when just one or two states have been used to test the action
of the quantum channel no regularization is needed to estimate the channel.\newline
{\bf (1)} When we use only a single test state (e.g., a pure state) the resulting estimated map
is a contraction of the Bloch sphere into a line connecting
two mutually orthogonal states. If we use the data $\varrho_z\to\varrho_z^\prime$,
then this line is given by the points $(0,0,\pm 1)$,
i.e. the Bloch sphere is mapped into the $z$ axis. This map
can be understand as the ``classical'' logical NOT. \newline
{\bf (2)} In the case of two test states, we obtain a unitary rotation.
If we use $\varrho_z\to\varrho_z^\prime$
and $\varrho_y\to\varrho_y^\prime$, then the result is
a rotation by the angle $\pi$  around the $x$ axis, i.e. the $\sigma_x$
operation. This operation is usually referred to as the ``quantum'' (unitary) logical NOT
(see Fig.\ref{incomplete}).

When three test states are used to determine the action of the universal NOT gate
we face a serious difficulty: Let us consider transformations of three test states
$\varrho_{x,y,z}\to\varrho_{x,y,z}^\prime$. In this case no completely
positive map exists. Therefore, a regularization of
output states is required - in particular, we have to determine a minimal amount of noise that is ``included''
in the output states so that the channel estimation with these regularized (noisy) output states will result
in a CP map. It turns out that the amount of noise corresponds to the value $k=1/3$. Therefore the transformation
that is closest to the universal NOT gate and is CP has the form
\be
\varrho_j^\prime=\frac{1}{3}\varrho_j^T+
\frac{2}{3}\frac{1}{2}\openone\, .
\ee
We can conclude that already for three test states we find that the reconstructed map that
is CP is  the optimal universal NOT operation \cite{werner}. In Fig.\ref{incomplete}
we  see how the estimation of the map changes with the number $n$ of used test states.

\begin{figure}
\begin{center}
\includegraphics[width=8cm]{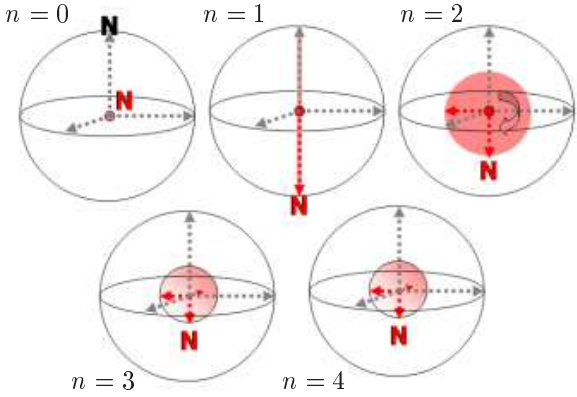}
\caption{
The figure represents  how our
knowledge about the channel (the universal NOT gate)
is improved with increasing the number of test states when the second reconstruction strategy
is used.
If no measurement has been performed then the best possible
estimation of the channel is the contraction of the Bloch sphere
into its center ($n=0$). When a single test state
$\varrho_z\to\varrho_z^\prime$ is used  the optimal estimation of the transformation
is a contraction of the whole Bloch sphere into a line connecting the ``north'' and
``south'' poles of the Bloch sphere (i.e. the points  $(0,0,\pm 1)$, respectively). This
contraction is represented by the figure $n=1$ in the picture.
When two test states ($n=2$) are used,
 i.e. $\varrho_z\to\varrho_z^\prime$
and $\varrho_y\to\varrho_y^\prime$, then the result of the estimation of the action of the quantum
channel is described by the transformation (19). Consequently, the map is unital and has the diagonal form
with
$\lambda_1=0,\lambda_2=\lambda_3=-1$. Unfortunately this map is not completely positive. The regularization
parameter  $k=1/2$ is used to correct this map.
Finally, when three ($n=3$) or four ($n=4$) test states are
used to determine the action of the quantum channel prescribed by the universal NOT transformation, the
best estimation of the channel is the optimal universal NOT gate.
}
\label{incomplete2}
\end{center}
\end{figure}

In what follows we will study the second
strategy for a reconstruction (estimation) of the best possible
approximation of the universal NOT gate.
Instead of searching for a completely positive map by ``admixing''
the smallest possible amount of noise into the outputs of the test states, let us assume
that all states except the test states are mapped into the total mixture \footnote{More specifically,
states that are complemented to a linear span of the test states.}.
Using this approach we can be sure that the resulting map is
positive, but not completely positive. Therefore we have to use a regularization of the map.
Obviously, in this case the resulting map will not explicitly satisfy conditions imposed by the
transformation of test states, $\varrho_j\to\varrho_j^\prime$, i.e. with the
given data. But this has to be expected since it is the only way how to impose the CP condition
on the reconstructed map. Using this type of regularization of the map we arrive at the same estimations
as in the case of previous strategy.
However, for $n=2$ (two test states) the situation is different. In this case,
the reconstructed map acts as follows
\be
\vec{r}=(x,y,z)\to\vec{r}^\prime=(0,-y,-z)\; .
\ee
That is, the states of the form $\varrho=\frac{1}{2}(\openone+x\sigma_x)$
are transformed into the total mixture.
Consequently, the map is unital and has a diagonal form with
$\lambda_1=0,\lambda_2=\lambda_3=-1$, which is not compatible with
the complete positivity  (see Fig.\ref{unital_cp}). The parameter $k=1/2$
can be used to correct this map. This value of $k$ can be derived
from the inequalities $|\lambda_1\pm\lambda_2|\le|1\pm\lambda_3|$.
One can observe this result in  Fig.\ref{unital_cp} as the closest point
from the tetrahedron to the point $(0,-1,-1)$.
The whole incomplete reconstruction using this method
is depicted in Fig.\ref{incomplete2}. In Tab.~1
we  present  diagonal elements (all other matrix elements are equal to zero) of matrices corresponding to reconstructed maps
based on the results of measurement of $n$ test states. We present results for both reconstruction methods.
\begin{table}
\begin{tabular}{|l||c|c|}
\hline
 & 1st method & 2nd method \\
 \hline
$\cE_0$ & $\{1,0,0,0\}$ & \{1,0,0,0\} \\
$\cE_1$ & $\{1,0,0,1\}$ & $\{1,0,0,1\}$\\
$\cE_2$ & $\{1,1,-1,-1\}$ & $\{1,0,-1/2,-1/2\}$ \\
$\cE_3$ & $\{1,-1/3,-1/3,-1/3\}$ & $\{1,-1/3,-1/3,-1/3\}$ \\
$\cE_4$ & $\{1,-1/3,-1/3,-1/3\}$ & $\{1,-1/3,-1/3,-1/3\}$\\
\hline
\end{tabular}
\caption{The two reconstruction scenarios for an un-physical universal NOT gate
result in a sequence of maps depending on the number of test states. The subscript in the description of a given map $\cE_j$
corresponds to the number of test states that have been used in the reconstruction. From the
table it is clear, that both methods give us the same process estimation except for the
case when two test states have been used.}
\end{table}

\section{Conclusion}

Reconstruction of quantum maps is a challenging problem
motivated mainly by experimental realizations of quantum
gates. These gates have to be tested thoroughly in order
to use them for any quantum computation. Another important application
of a channel reconstruction is in quantum communication when characteristics
of a quantum channel have to be determined from a limited set of tests performed
on the channel.

In the present paper we have discussed some strategies how to estimate
quantum channels when only incomplete and/or incompatible data from
measurements is available. The incompatibility of available data results in estimated
maps that are not completely positive. In this case a regularization of
the reconstruction procedure is required in order to recover a physical
(CP) map.

The regularization method represents  a correction
of insufficient statistics. It uses a single parameter $k$ that can be understood as an addition of a white
noise into our data. In principle, we can face two situations: Either
reconstructions of all test states at the output of the channel correspond to proper quantum states, or there exist
some outcomes that are not proper quantum states. In this case each of these states
can be corrected by adding some noise (using the multiplicative factor $k$ as discussed above).
We have to keep in mind that the parameter $k$
can be state dependent. Therefore, we have to choose
the smallest value of $k$ in order to correct the map. Once all the outputs (as estimated from
the measured data) are proper physical states we can start to reconstruct the map itself.
In spite of the fact that all the test states at the output are proper physical states
the map that is estimated on the basis of these states
may be not completely positive. In this case
we have to search for the largest $k$, for which the
corrected map $\cE_c=k\cE+(1-k)\cA$ is completely positive.
We have shown that at least for unital maps the value
$k=1/3$ always regularizes the estimated map.

In order to illustrate our methods we have
used the data generated by an unphysical map - the universal NOT gate.
We have shown that using both approaches the result
is the same and the reconstructed map is the best (optimal) approximation
of the NOT operation \cite{werner}.

\section*{Acknowledgment}
This was work supported in part by  the European
Union projects QUPRODIS (IST-2001-38878)  and
CONQUEST (MRTN-CT-2003-505089). M.Z. would like to thank for a partial support via
the grant 201/04/1153 of the GACR.


\end{multicols}


\end{document}